# A featureless transmission spectrum for the Neptune-mass exoplanet GJ 436b

Heather A. Knutson[1], Björn Benneke[1,2], Drake Deming[3], & Derek Homeier[4]

[1]Division of Geological and Planetary Sciences, California Institute of Technology, Pasadena, CA 91125, USA. [2]Department of Earth, Atmospheric, and Planetary Sciences, Massachusetts Institute of Technology, Cambridge, MA 02139, USA. [3]Department of Astronomy, University of Maryland, College Park, MD 20742, USA. [4]Centre de Recherche Astrophysique de Lyon, 69364 Lyon, France.

**GJ 436b is a warm (approximately 800 K) extrasolar planet that periodically eclipses its low-mass (0.5 $M_{Sun}$) host star, and is one of the few Neptune-mass planets that is amenable to detailed characterization. Previous observations[1,2,3] have indicated that its atmosphere has a methane-to-CO ratio that is $10^5$ times smaller than predicted by models for hydrogen-dominated atmospheres at these temperatures[4,5]. A recent study proposed that this unusual chemistry could be explained if the planet's atmosphere is significantly enhanced in elements heavier than H and He[6]. In this study we present complementary observations of GJ 436b's atmosphere obtained during transit. Our observations indicate that the planet's transmission spectrum is effectively featureless, ruling out cloud-free, hydrogen-dominated atmosphere models with a significance of 48σ. The measured spectrum is consistent with either a high cloud or haze layer located at a pressure of approximately 1 mbar or with a relatively hydrogen-poor (3% H/He mass fraction) atmospheric composition[7,8,9].**

We observed four transits of the Neptune-mass planet GJ 436b on UT Oct 26, Nov 29, and Dec 10 2012, and Jan 2 2013 using the red grism (1.2-1.6 μm) on the *Hubble Space Telescope (HST)* Wide Field Camera 3 instrument. These data were obtained in a new scanning mode[10,11] with a scan rate of 0.99"/s, which allowed us to achieve approximately a factor of twenty improvement in the orbit-averaged efficiency as compared to staring-mode observations[12]. Each visit spanned four *HST* orbits with an integration time of 7.6 s per exposure. We extract the spectra from the raw images using the template-fitting technique described in a previous study[11], and provide additional details of our reduction pipeline in the Methods section.

We fit the four wavelength-integrated (white-light) transit curves simultaneously[13] while accounting for detector effects (see discussion in Supplemental Methods) in order to determine values for the planet's orbital inclination $i$, the planet-star radius ratio $R_p/R_*$, the ratio of the planet's semi-major axis $a$ to the stellar radius $R_*$, and the center-of-transit time $T_c$. We set the uncertainties on each measurement equal to the standard deviation of the residuals from our best-fit solution for that visit and evaluate the uncertainties on our best-fit parameters using the covariance matrix from our Levenberg-Marquardt least-squares minimization, a Markov Chain Monte Carlo analysis, and a residual permutation technique that better accounts for the presence of time-correlated noise in the data[14,15]. The residual permutation approach results in uncertainties that are a factor of 1.5-2 larger than both the covariance matrix and the MCMC errors for all of our fit parameters, and we take those as our final errors.

Our best-fit parameters for the white-light transits are given in Table 1, and the normalized transit light curves are shown in Fig. 1. We see no evidence for transit depth variations comparable to those reported with the *Spitzer* data, and we do not detect any star spot occultations in our transit light curves comparable to those observed for HD 189733b[16,17]. We next determine the differential wavelength-dependent transit depths in twenty-eight bins spanning wavelengths between 1.14-1.65 µm as described in Fig. 2 and the Methods section. The resulting transmission spectrum is shown in Fig. 2, with error bars that include both the uncertainties in the measured transit depth and in the stellar limb-darkening models.

We interpret the transmission spectrum using a variation of the Bayesian atmospheric retrieval framework described in a previous study[18]; we provide a summary of this approach in the Methods section. We find that we obtain the best match to our data using models with either high altitude clouds or a relatively hydrogen-poor atmosphere with a reduced scale height and correspondingly small absorption features. At low metallicities our model requires a haze or cloud layer at pressures below 10 mbars, as the large scale height of these models otherwise leads to strong spectral signatures from molecular absorption (e.g., $H_2O$, $CH_4$, CO, or $CO_2$). At higher metallicities the scale height of the atmosphere is reduced, and no clouds are needed in order to produce an effectively flat transmission spectrum. Our conclusions are similar to those obtained for the transiting super-Earth GJ 1214b[12], although in this case new upper limits on the planet's transmission spectrum indicate that it must have a high cloud layer even if the atmosphere is metal-rich[19]. GJ 436b is four times more massive with a nearly identical average density and therefore seems a less obvious candidate for a hydrogen-poor atmosphere. The 68% ($1\sigma$) Bayesian credible region extends along a curve from hydrogen-dominated models with a high-altitude cloud layer between 0.01 and 4 mbars to high-metallicity models that may or may not contain clouds.

For solar composition atmospheres, a cloud or haze layer at 1 mbar that is optically thick for slant viewing geometries represents the best fit to the data. Zinc sulfide (ZnS) and potassium chloride (KCl) are both plausible cloud candidates, as the condensation curves of these substances can readily cross the pressure-temperature profile at the mbar level in GJ 436b's atmosphere[6,20]. A recent study[20] of the super-Earth GJ 1214b, which is also a good candidate for these cloud species, indicates that a solar composition atmosphere would not have sufficient amounts of condensable material to form optically thick clouds. If the atmospheric metallicity (defined as the abundances of elements heavier than H and He) is enhanced above this level, then such clouds could easily explain this planet's flat transmission spectrum[20]. Alternatively, photochemical haze production could lead to an opaque cloud layer at the mbar level, although these models also likely require an enhanced atmospheric metallicity.

Previous studies have placed constraints on the possible bulk compositions for GJ 436b using the planet's measured mass and radius, the estimated age of the system, and models of planet formation and migration[7,8,9,21,22,23]. A recent survey of the published literature for this planet indicated that current models are consistent with bulk metallicities between 230-2000 times solar depending on the assumed ratio of rock to ice, the distribution of metals between the core and the envelope, the interior temperature of the planet, and other related factors[6]. This corresponds to a H/He mass fraction of approximately 3-22%. Based on this analysis we conclude that an atmospheric metallicity of 1900 times solar is consistent with current constraints for the planet's bulk composition, although it is very close to the upper end of this range. Mass loss appears to be minimal for GJ 436b under present-day conditions[24,25,26], but it is possible that the higher UV and X-ray fluxes expected for young stars could have resulted in the loss of some atmospheric hydrogen very early in the planet's history[27,28]. Although a recent study[29] argued that such mass loss is unlikely to result in significant depletion of hydrogen relative to other elements, additional modeling work is still needed in order to provide a more definitive resolution to this question.

There are several potential avenues to distinguish between cloudy and high atmospheric metallicity scenarios for GJ 436b's atmosphere. An unambiguous solution would be to obtain more precise, moderate-resolution transmission spectroscopy[30] capable of detecting near-infrared absorption features and directly constraining the mean molecular weight. We emphasize, however, that the apparent variations in the planet's measured transit depth from one epoch to the next (see Methods) make simultaneous measurements essential for robust constraints on this planet's transmission spectrum. A hydrogen-dominated atmosphere with a high cloud or haze layer should exhibit attenuated water absorption features, which could be distinguished from the intrinsically weaker features of high metallicity atmospheres based on the steepness of the wings of the absorption lines. Similarly, a detection of the Rayleigh scattering slope in the planet's visible-light transmission spectrum could directly constrain both the mean molecular weight and the amount of spectrally inactive gas (i.e., H and He or haze particles) present in the atmosphere[20,30]. Alternatively, one could differentiate between hydrogen-dominated and high metallicity atmospheres by measuring the relative abundances of CO, $CO_2$, methane, and water. These relative abundances could then be compared to different chemical models for GJ 436b's atmosphere that might rule out the presence of significant molecular hydrogen. Improved constraints on the atmospheric chemistry from secondary eclipse spectroscopy[1,2], which is less sensitive to high-altitude clouds and stellar activity, could also help to restrict the range of plausible atmospheric compositions in the limit of a well-mixed atmosphere (i.e., no significant compositional gradients between the day and night sides). Lastly, improved estimates for the stellar mass and radius would help to reduce the uncertainties in the corresponding planetary values and hence better constrain its mean density.

**Methods Summary**
We calculate this transmission spectrum as follows: first we determine the difference between our extracted spectrum and a best-fit template spectrum at each pixel position and create a time series of the residuals. We then fit this time series with a model consisting of the difference between the white-light transit curve and a transit light curve with a freely varying planet-star radius ratio. We also include a linear function of time to account for the first order of any remaining instrumental trends. We compare the errors on the planet-star radius ratio from the Levenberg-Marquardt covariance matrix and the residual permutation method and take the larger of the two as our final uncertainty; they typically agree to within 10%. We then average the planet-star radius ratios in four-pixel-wide segments to create our final transmission spectrum for each visit, where we select our wavelength range to exclude the low-illumination regions at the edges of the spectrum. We calculate uncertainties on each bin as the average of the errors for the four individual radius ratios in order to account for the four-pixel-wide Gaussian smoothing function we applied to the raw spectra before fitting the template spectrum. We combine the data from our four visits by taking the error-weighted mean of the transit depths in each wavelength bin.

**Acknowledgements.** We thank Peter McCullough for his assistance in the planning and executing of these observations. We are also grateful to Julianne Moses, Michael Line, and Nadine Nettelmann for conversations on the nature of high-metallicity atmospheres as well as discussions of specific interior and atmosphere models for GJ 436b. D.H. has received support from the European Research Council under the European Community's Seventh Framework Programme (FP7/2007-2013 Grant Agreement no. 247060).

**Author Contributions.** H.A.K. carried out the data analysis for this project with input from D.D. B. B. provided the planetary atmosphere models and accompanying fits, while D. H. supplied the PHOENIX atmosphere models used to calculate the stellar limb-darkening coefficients.

**Author Information.** Reprints and permissions information is available at www.nature.com/reprints. The authors have no competing financial interests to report. Correspondence and requests for materials should be addressed to hknutson@caltech.edu.

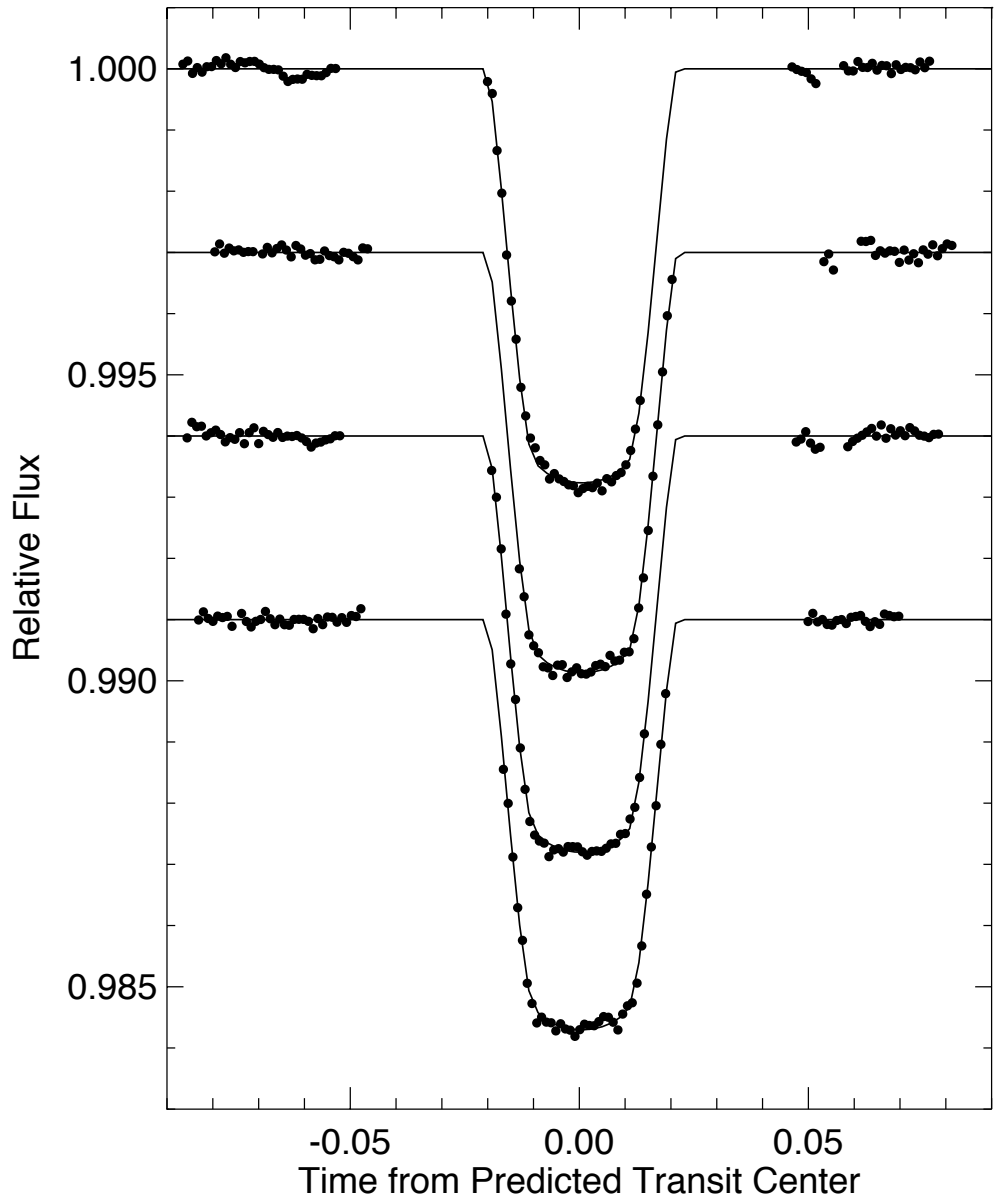

**Figure 1. White-light transit curves for the four individual visits.** Data are vertically offset for clarity. Transits were observed on the following dates (from top to bottom): UT Oct 26, Nov 29, and Dec 10 2012, and Jan 2 2013. Normalized data with the first orbit trimmed and instrumental effects removed are shown as black filled circles. Best-fit model transit light curves are shown as black lines. The data consist of three spacecraft orbits with durations of approximately 1.5 hours each; there is a gap during each orbit where the spacecraft passes behind the Earth and the target is no longer visible.

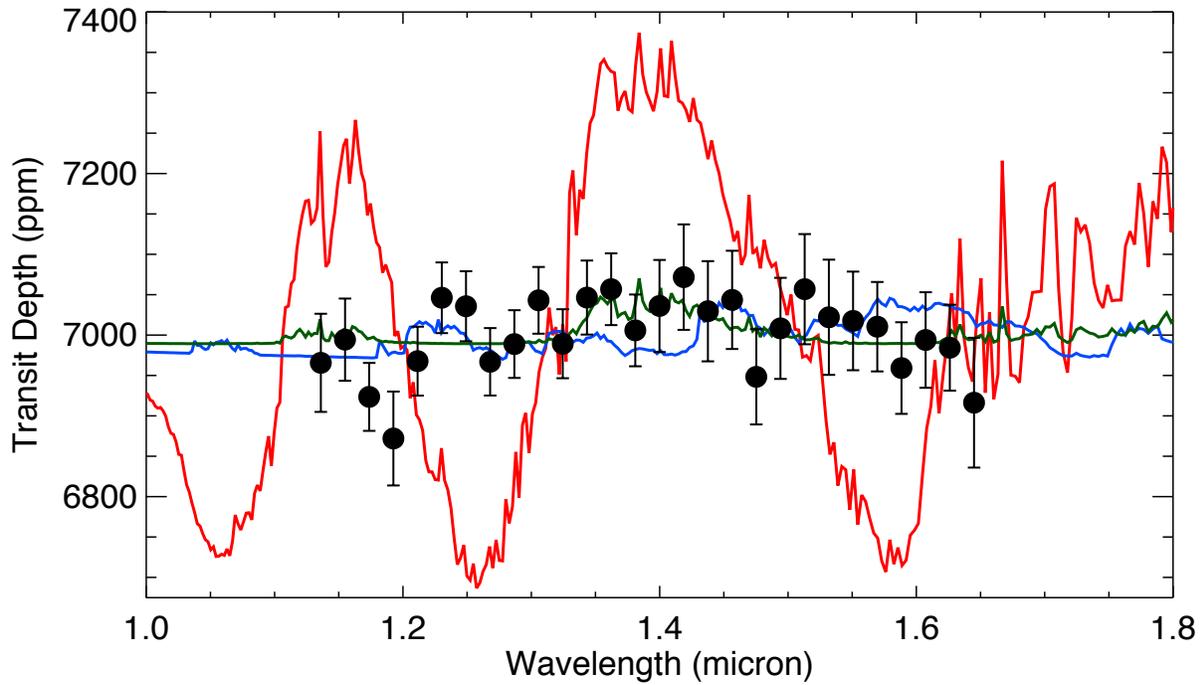

**Figure 2. Averaged transmission spectrum for GJ 436b.** Black filled circles indicate the error-weighted mean transit depth in each bandpass, with the plotted uncertainties calculated as the sum in quadrature of the 1σ standard deviation measurement errors and the systematic uncertainties from stellar limb-darkening models. We show three models for comparison, including a solar-metallicity cloud-free model (red line), a hydrogen-poor 1900 times solar model (blue line), and a solar metallicity model with optically thick clouds at 1 mbar (green line).

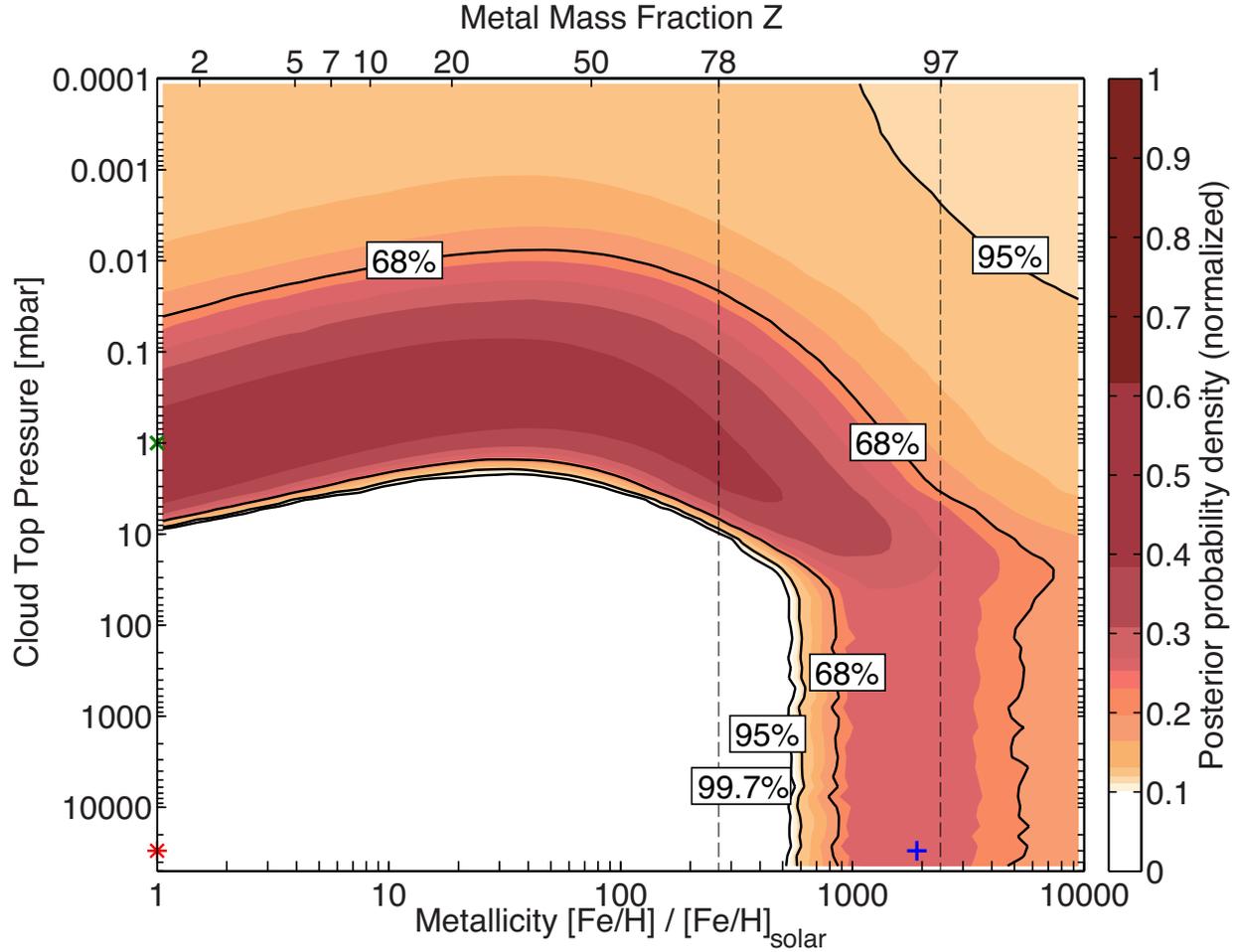

**Figure 3. Joint constraints on cloud top pressure versus atmospheric metallicity.** The colored shading indicates the normalized probability density as a function of metallicity (defined as the relative abundance of elements heavier than H and He) and cloud top pressure derived using our Bayesian atmosphere retrieval framework[18,30]. The black contours show the 68%, 95%, and 99.7% Bayesian credible regions. We rule out a hydrogen-dominated, cloud-free, solar-metallicity atmosphere with a significance of 48σ. Color-matching markers indicate the three models plotted in Figure 2, and vertical dashed lines indicate constraints on the planet's composition from measurements of its average density.

**Table 1. Best-fit 1σ transit parameters from white light curves.** We find that our estimates for the orbital inclination $i$ and the ratio of the semi-major axis $a$ to the stellar radius $R_*$ are consistent at the 3σ level with our previously published 8 μm *Spitzer* transit observations[3], and that our individual values for the planet-star radius ratio $R_p/R_*$ are mutually consistent at the 3σ level. We derive a new transit ephemeris of with a center-of-transit time $T_0$=2456295.431924(45) BJD$_{TDB}$ and an orbital period $P$=2.64389782(8)[a] days by combining our data with previous studies[3]; our new data extend the current baseline for this object by almost four years.

| Visit Date | $R_p/R_*$ | $T_C$ (BJD$_{TDB}$[b]) |
|---|---|---|
| UT Oct 26 2012 | 0.08349(33) | 2456226.69131(11) |
| UT Nov 29 2012 | 0.08413(26) | 2456261.06211(10) |
| UT Dec 10 2012 | 0.08372(31) | 2456271.63758(10) |
| UT Jan 2 2013 | 0.08310(27) | 2456295.43270(7) |
| **Averaged Values** | | |
| Inclination $i$ (º) | 86.774(30) | |
| $a/R_*$ | 14.41(10) | |

[a]Errors are given in parenthesis after the best-fit value, with the number N of significant digits in the quoted errors corresponding to the last N significant digits of the best-fit value. For example, the stated orbital period corresponds to a value of 2.64389785 ± 0.00000008 day using the standard notation.

[b]Barycentric Julian date for the measured center-of-transit times. To convert from TDB to UTC time standards, simply subtract 67.18 s from the reported center-of-transit times.

# Methods
## 1. Spectral extraction

We follow the method described in a previous study[11] and summarize the main steps here for reference. We obtain our data using the 256x256 pixel subarray and the SPARS10 readout mode with two samples, which has an effective integration time of 7.62 s. These data are available for download from the MAST archive as part of proposal number 11622. We extract our spectra from the raw _ima.fits image files (see Extended Data Figure 1 for a representative image), as the method for calculating the fluxes for the _flt.fits files does not work for data obtained in drift scan mode. The _ima.fits files were processed using either version 2.7 (all 2012 transits) or version 3.0 (Jan. 2013 transit) of the CALWF3 pipeline, which applies a standard set of calibrations including dark subtraction, linearity correction, cosmic ray rejection, and a conversion from raw counts to flux units as described in section 3.2.3 of the WFC3 Data Handbook.

The _ima.fits files retrieved through MAST contain an array of three images produced by the sample-up-the-ramp readout. These images were taken 0, 0.28, and 7.62 s. after the start of the exposure; in our subsequent descriptions we will refer to them as images A, B, and C, respectively. We convert each image from units of $e^- s^{-1}$ to $e^-$ using the appropriate integration times and difference each pair of sampled images (e.g., B-A and C-B). We then trim out a central region encompassing the location of the spectrum in each differenced image and add the trimmed differenced images together (e.g., (B-A) + (C-B)) to create our final science image. Because we use a different sub-aperture for each differenced image, our final combined (B-A) + (C-B) science image is not simply equal to the (C-A) image. Previous studies[11] used this differenced image approach to minimize contamination from the sky background in the scanned images, therefore avoiding the need for a separate background subtraction step. Although we adopt the same approach in this study, our mask is only excluding the sky background from the first 0.28 s of the 7.62 s integration and we therefore include a separate sky subtraction step later in our analysis. We find that for our data this image differencing approach gives results that are identical to the case where we simply use the third (C) sample-up-the-ramp image as our science image with no subtraction or masking.

We next select a sub-aperture centered on the position of the stellar spectrum in our science image with dimensions of 160 pixels in the x (dispersion) direction and 71 pixels in the y (cross-dispersion) direction. We use a fixed aperture position for each visit, and estimate the position of the star using an acquisition image obtained at the start of each visit. Unlike previous studies[11], we find that using a narrow aperture that cuts off at half the maximum flux produces an increased scatter in our white-light photometry; this may be due to the larger orbit-to-orbit position drift in our images as compared to HD 204958b. In this case we obtain optimal results with an aperture that extends out to the wings of the point spread function in both dimensions.

We apply a color-dependent flat-field correction and calculate wavelength solutions for our differenced images using coefficients adapted from the standard STScI pipeline as described in other studies[11,31]. We then apply a filter to remove bad pixels and cosmic ray hits by first dividing each row of the individual subarrays by the total flux in that row (this corrects for the uneven scan rate in the y direction) and then iteratively flagging $8\sigma$ and then $4\sigma$ outliers in the time series at each pixel position using a moving median filter with a width of five pixels (i.e., we calculate the median flux value at that pixel position starting from two images before and ending two images after our science image). We replace flagged pixels with the value of the moving median filter at that position, then multiply each row by the original flux total from that row to restore the initial subarray with bad pixels removed. Approximately 0.04-0.06% of the pixels within the subarray aperture are flagged as bad in our four visits. We then sum in the y (cross-dispersion) direction to create a one-dimensional spectrum from each image.

We calculate the MJD mid-exposure times corresponding to each spectrum from the headers of the .flt files, and convert these times to $BJD_{TDB}$ using publically available routines[32]. The median sky background in our _ima.fits images is approximately 0.1% of the total flux when we sum over the spectrum. We see no evidence for any wavelength or time dependence in the background flux, and so simply subtract the median background level from each visit.

As noted in a previous study[11], the WFC3 spectra are undersampled and this can create problems when fitting templates with slightly offset positions in the dispersion direction. We mitigate this issue by convolving all of our 1D spectra with a Gaussian function with a full width at half max (FWHM) of 4 pixels; this modestly degrades the wavelength resolution of our spectra, but the loss is negligible since we ultimately bin our transmission spectrum in four-pixel-wide bins. We next create a template spectrum for each visit by averaging ten spectra immediately before and immediately after the transit. We fit the template spectrum to the central 112 pixels of the individual spectra from each visit, allowing the template amplitude to vary freely and the relative positions to shift by increments of $1/1000^{th}$ of a pixel.

## 2. Transit Fits
The template-fitting technique results in two kinds of data products: first, a white-light curve for each transit calculated from the best-fit amplitude for the template spectrum at each time step, and second, a set of wavelength-dependent time series calculated from the difference of the best-fit template spectra and the measured spectra at each pixel position. This method is designed to remove common-mode white-light instrumental effects from the differenced spectra without the need to assume a functional form for these effects, resulting in lower noise levels in the final transmission spectrum than other commonly used approaches[12,32,33].

## 2.1 White Light Transit Fit
Our fits to the white-light transit photometry (Fig. 1 and Extended Data Fig. 2 and 3) include a linear function of time and a linear plus exponential function of orbital phase (five free parameters) in order to describe the behavior of the instrument. We assume that $i$ and $a/R_*$ are the same in all visits, but allow $R_p/R_*$, $T_c$, and the instrumental terms to vary individually. We trim the data from the first orbit in our fits to the white-light data, as these data display larger-than-usual instrumental effects due to settling at the new pointing. We keep this first orbit when we fit the differential transmission spectra (Extended Data Fig. 4) as there does not appear to be any evidence for color-dependent instrumental effects at the start of each light curve and this gives us a longer baseline for our residual permutation error estimation. We set the uncertainties on each white-light measurement equal to the standard deviation of the residuals from our best-fit solution for that visit, which are equal to $[10.0, 9.1, 8.9, 7.8] \times 10^{-5}$ for the Oct., Nov., Dec., and Jan. visits, respectively. These residuals are a factor of 1.2-1.5 times higher than the white-light photon noise limit of $6.4 \times 10^{-5}$, reflecting the uncorrected instrumental effects visible in Extended

Data Fig. 3. The $\chi^2$ value for our simultaneous fit to all four white-light transit curves is 356.7, with a total of 360 points in our fit and 26 free parameters. We also compare the rms of the residuals in our four-pixel bands to the photon noise limit in those bands and find median values ranging between 1.03-1.07 times the photon noise limit for our four individual transit observations. We calculate our errors on the wavelength-dependent transit depths

We show our best-fit transit times in comparison to previously published values in Extended Data Fig. 5. Although we do not expect the planet radius to vary in time, previous studies have reported variations in the measured transit depths at different epochs, which could be caused by the occultation of bright or dark regions on the stellar surface[3,34]. We show the best-fit transit depths from our four white-light curves in comparison to these previous studies in Extended Data Fig. 6. In Extended Data Fig. 7 we plot the Ca II H & K activity index for this star as a function of time; although sampling is poor at the epoch of our HST observations, the star appears to have an average-to-low activity level at this time. This may explain the relatively small scatter in our measured transit depths over the two months spanned by our four transit observations as compared to prior *Spitzer* observations. We calculate a reduced $\chi^2$ value of 2.7 for our four transit depths as compared to the averaged value, suggesting that stellar activity may still be contributing some extra variability.

## 2.2 Wavelength-Dependent Transit Fit
We calculate the transmission spectrum as follows: first we determine the difference between our extracted spectrum and a best-fit template spectrum at each pixel position and create a time series of the residuals. We then fit this time series with a model consisting of the difference between the white-light transit curve and a transit light curve with a freely varying planet-star radius ratio. We also include a linear function of time to account for the first order of any remaining instrumental trends. We compare the errors on the planet-star radius ratio from the Levenberg-Marquardt covariance matrix and the residual permutation method and take the larger of the two as our final uncertainty; they typically agree to within 10%. We then average the planet-star radius ratios in four-pixel-wide segments to create our final transmission spectrum for each visit (Extended Data Fig. 4), where we select our wavelength range to exclude the low-illumination regions at the edges of the spectrum. We calculate uncertainties on each bin as the average of the errors for the four individual radius ratios in order to account for the four-pixel-wide Gaussian smoothing function we applied to the raw spectra before fitting the template spectrum. We combine the data from our four visits by taking the error-weighted mean of the transit depths in each wavelength bin (Fig. 2).

## 2.3 Limb-Darkening Models
We compare results for both our white-light fits and our differential transmission spectra using fixed four-parameter nonlinear limb-darkening coefficients calculated from a PHOENIX stellar atmosphere model[35]. We first calculate the average center-to-limb intensity profile for the nominal wavelength range of each individual pixel, then convolve the resulting model spectrum at each radial position on the star with a four-pixel wide Gaussian function in wavelength space in order to account for the smoothing applied to our measured spectra. We then fit the smoothed intensity profiles at each pixel position with a four-parameter nonlinear limb-darkening profile[36] and use those limb-darkening coefficients to calculate our transit light curves.

A recent study[37] estimated an effective temperature of 3416 ± 54 K for GJ 436 based on new interferometric radius measurements. We therefore consider four different stellar atmosphere models with effective temperatures ranging between 3350 K and 3500 K. We show plots of the disk-integrated spectra for the hottest and coldest models compared to the measured spectra for each visit in the WFC3 band in Extended Data Fig. 8. Our choice of limb-darkening model has a relatively small effect on the overall shape of our measured transmission spectrum, and we quantify this effect as a systematic error term in Extended Data Table 1. We estimate the contribution of the limb-darkening model errors by calculating the change in the measured transit depth in a given band over a range of 3350-3450 K in the stellar effective temperature used for the limb-darkening models. We then add these errors in quadrature to the measurement errors when comparing our results to model transmission spectra, and show the combined errors in Fig. 2.

We also compare the fits to the white light curves using different stellar atmosphere models and find a $\chi^2$ value of 357.2 for the 3350 K model, 356.7 for the 3400 K model, 357.2 for the 3450 K model, and 357.6 for the 3500 K model, an effectively negligible difference. Without a strong preference for one model over the other, we elect to use the 3400 K model in our final analysis for consistency with the published temperature estimate[37]. We tried fits with a linear limb-darkening coefficient as a free parameter at each wavelength, where we constrained these coefficients to vary within the range spanned by the model coefficients for stellar effective temperatures between 3350-3450 K. We obtained a transmission spectrum that was consistent with our previous results, but with significantly larger uncertainties. This may be due to our choice of a linear parameterization for limb-darkening, which provides a quantifiably poorer fit to the white-light transit curves, or to weak constraints on the limb-darkening profile due to GJ 436b's near-grazing geometry (b=0.85; see our previous study of this planet[3] for a more detailed discussion of this geometry and its effect on our ability to empirically constrain limb-darkening profiles).

**4. Atmospheric Retrieval**

The observed transmission spectrum is interpreted using a variant of the atmospheric retrieval method described in previous studies[18,30]. The method used in this work combines a self-consistent, line-by-line atmospheric forward model and the nested sampling technique to efficiently compute the joint posterior probability distribution of the desired atmospheric parameters from the observed transmission spectrum. The main variation from the method described in our most recent paper[30] is that the analysis in this work employs our a-priori knowledge of chemistry to limit the range of atmospheric compositions to scenarios that are chemically plausible.

The goal of the retrieval analysis is to determine the range of metallicities (Fe/H) and cloud top pressures that are in agreement with the data (Fig. 3). Rather than fitting the data with unconstrained combinations of molecular abundances, however, we only compare the observations to atmospheres that are chemically plausible. Our approach is to compute the chemical equilibrium abundances and the temperature-pressure profiles self-consistently, while accounting for the uncertainties in the modeling of the methane abundance and the unknown Bond albedo through treating them as additional free parameters and marginalizing over them. In total, we perform a retrieval analysis in the five dimensional parameters space spanned by the metallicity, the cloud top pressure, the methane abundance relative to chemical equilibrium, the Bond albedo, and the reference planet-to-star radius ratio.

We introduce a free parameter for the methane abundance because the methane abundance has significant effect on the observed part of the transmission spectrum, but its abundance profile cannot be predicted reliably using self-consistent models. The dominant source of uncertainty in our estimates for the methane abundance for a given metallicity is introduced by our limited knowledge of the vertical pressure-temperature profile. The proximity of the expected temperature profile to the $CH_4$/CO transition[6] makes the methane abundance highly sensitive to the model assumptions on the vertical distribution of short wavelength absorbers, vertical energy transport, and day-night heat redistribution. Depending on whether the temperature in the photosphere is above or below the boundary where CO replaces $CH_4$ as the dominant carbon-bearing species, the methane abundance can vary by several orders of magnitude. Disequilibrium effects such as quenching and photochemistry at the upper end of the photosphere present an additional source of uncertainty. Our model determines the chemical composition of methane-reduced atmospheres by minimizing Gibb's free energy while simultaneously setting an upper limit on the methane abundance.

The other prominent absorber in the spectral range covered by our observations is water. We do not introduce an additional free parameter for water, however, because the water abundance in the photosphere can reliably be related to the metallicity of the atmosphere through chemical equilibrium calculations. Disequilibrium chemistry models for this planet[6] indicate that quenching and photochemical effects only affect the water abundance at pressures less than ~1 μbar, while our data are primarily sensitive to higher pressures. We include the Bond albedo as a free parameter to account for the uncertainty introduced by the effect of the albedo on the temperature pressure profile and therefore the atmospheric scale height. When calculating the significance with which the solar metallicity, cloud-free model is excluded we use the following definition[46]:

$$\text{Significance} = \frac{\chi^2_{obs} - \langle \chi^2 \rangle}{\sigma} = \frac{\chi^2_{obs} - \nu}{\sqrt{2\nu}}$$

where $\nu$ is the number of degrees of freedom in the fit. We calculate a value of 48σ from our final fits.

We also consider scenarios with either sub-solar (C/O = 0.3) or super-solar (C/O = 1.0) C to O ratios. We present the results from these retrievals in Extended Data Fig. 9. The oxygen-rich sub-solar C to O case produces results that are virtually identical to our solar C to O analysis in Fig. 3. For the carbon rich super-solar C to O case, the high metallicity (>1000x solar) cloud-free scenarios are excluded as they exhibit a CO absorption feature at 1.6 μm that appears to be inconsistent with our measured transmission spectrum. These same carbon-rich models also appear to allow a deeper (10x higher pressure) cloud deck for moderately metal-rich several hundred times solar scenarios. However, carbon-rich models with C/O > 0.8 do not provide a good fit to GJ 436b's dayside emission spectrum, as increasing the C to O ratio tends to increase the amount of methane in the atmosphere[6].

46. Gregory, P. Bayesian Logical Data Analysis for the Physical Sciences. Cambridge University Press: Cambridge (2005)

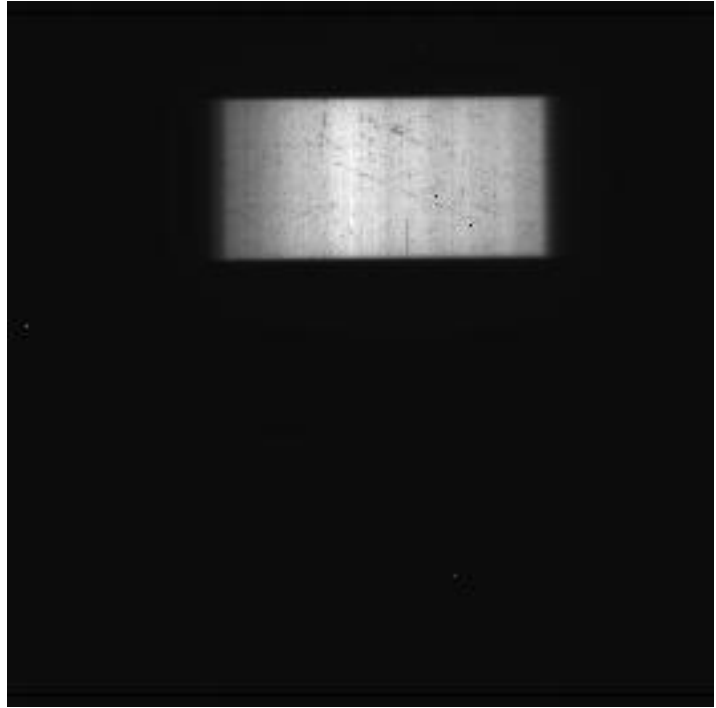

**Extended Data Figure 1.** Representative raw image from UT 2012 Nov 29 observation showing scanned spectrum.

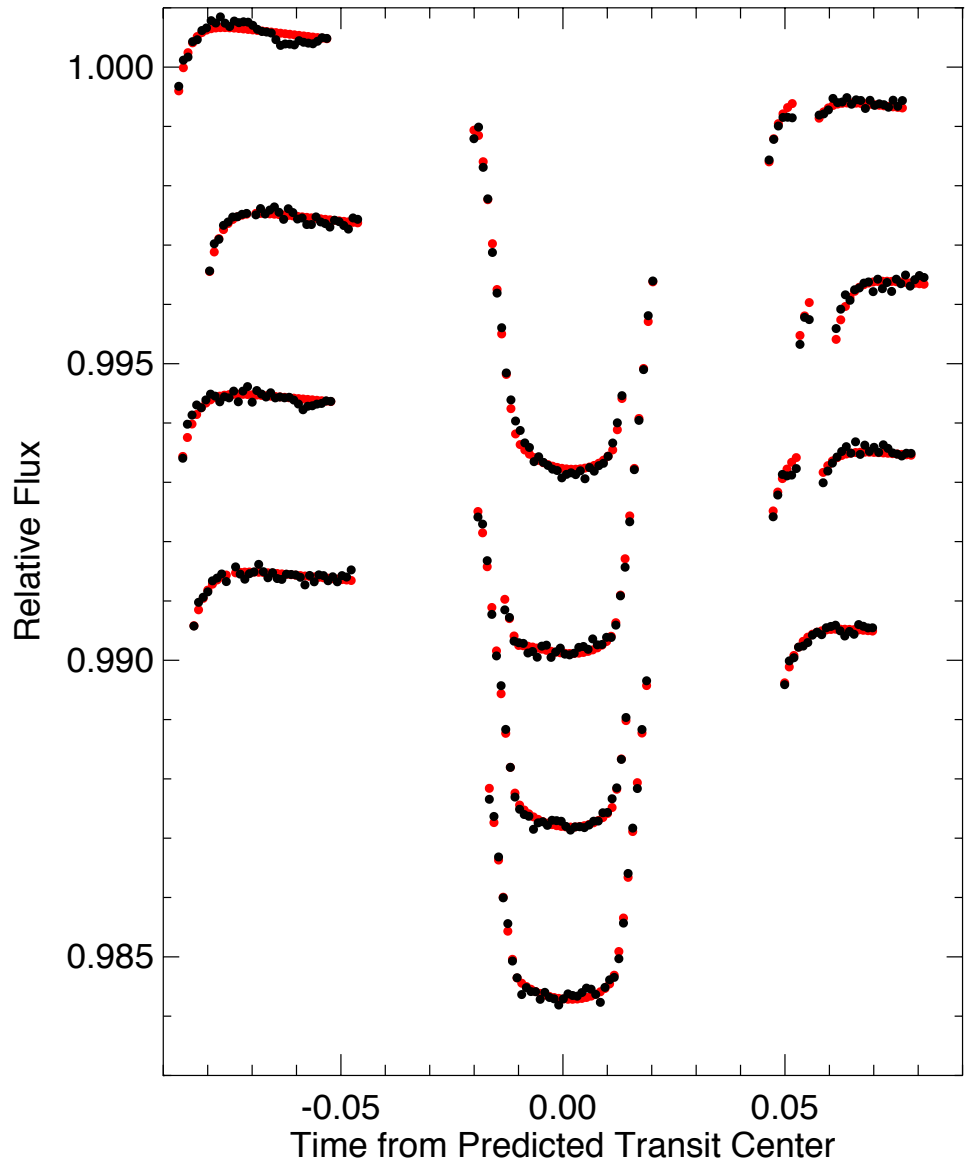

**Extended Data Figure 2. Raw white-light photometry for the four individual transits.** Data are vertically offset for clarity. Transits shown were obtained on the following dates (from top to bottom): UT Oct 26, Nov 29, and Dec 10 2012, and Jan 2 2013. The raw fluxes are shown as filled black circles, and the best-fit solutions for the instrumental effects and transit light curves are shown as filled red circles.

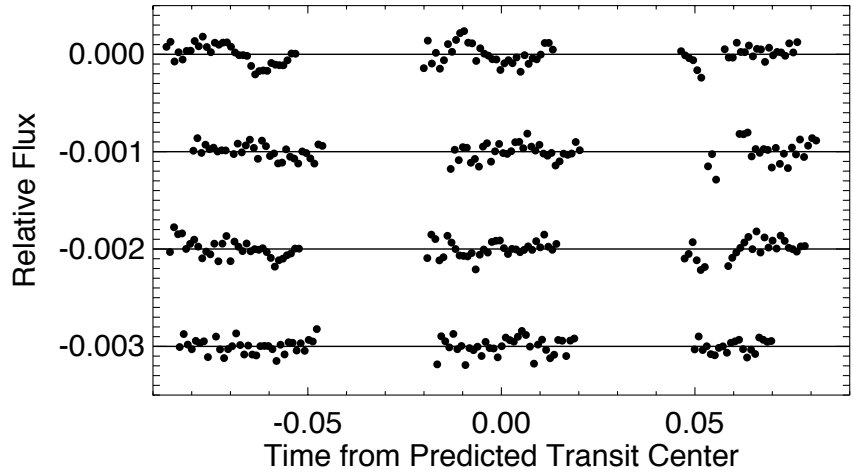

**Extended Data Figure 3. White-light residuals.** Data are vertically offset for clarity. Transit residuals shown were obtained on the following dates (from top to bottom): UT Oct 26, Nov 29, and Dec 10 2012, and Jan 2 2013. The difference between the white-light fluxes and best-fit model solutions are shown as filled black circles.

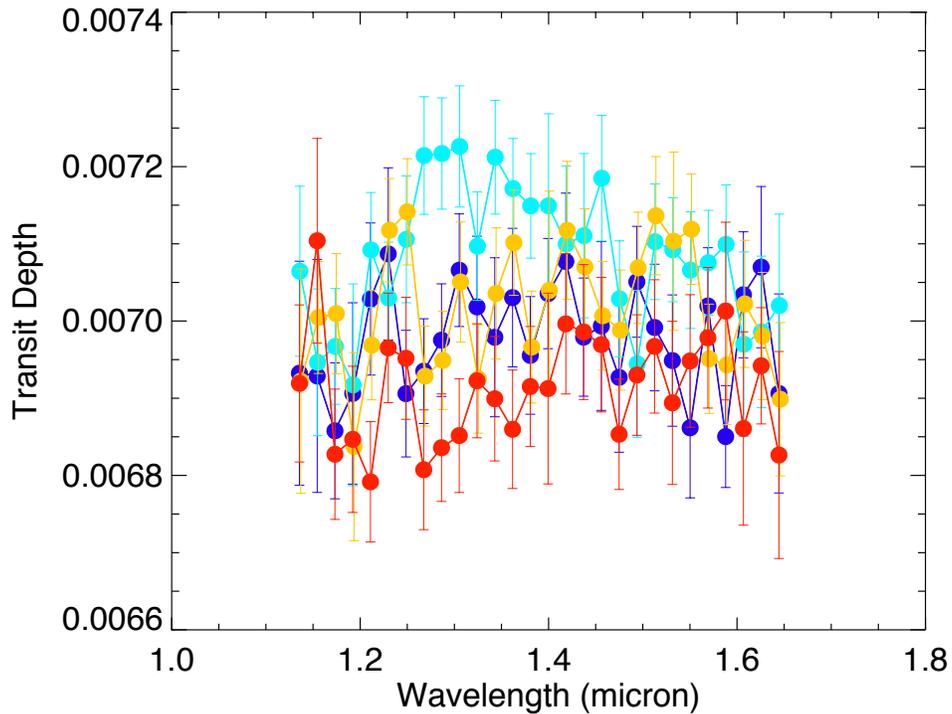

**Extended Data Figure 4. Individual transmission spectra for each of the four visits.** Transmission spectra are shown as filled circles, with colors indicating the date of the observations: UT Oct 26 (dark blue), Nov 29 (light blue), and Dec 10 2012 (yellow), and Jan 2 2013 (red). This plot shows the errors in the measured transit depths, but does not include the additional systematic errors from the limb-darkening models.

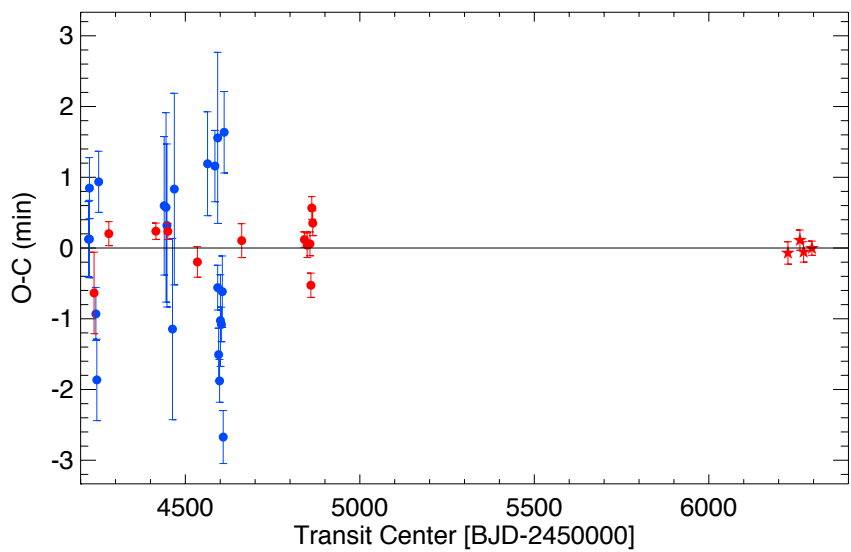

**Extended Data Figure 5. Observed minus calculated transit times using the new best-fit ephemeris.** The solid line denotes O-C equal to zero. Transit times from this paper are plotted as filled stars, while previously published observations are shown as filled circles. The color of the points denotes the wavelength of the observations (blue for visible, red for IR). Transits shown include all previously published observations for this planet[3,38-44]. Figure adapted from our previous study[3].

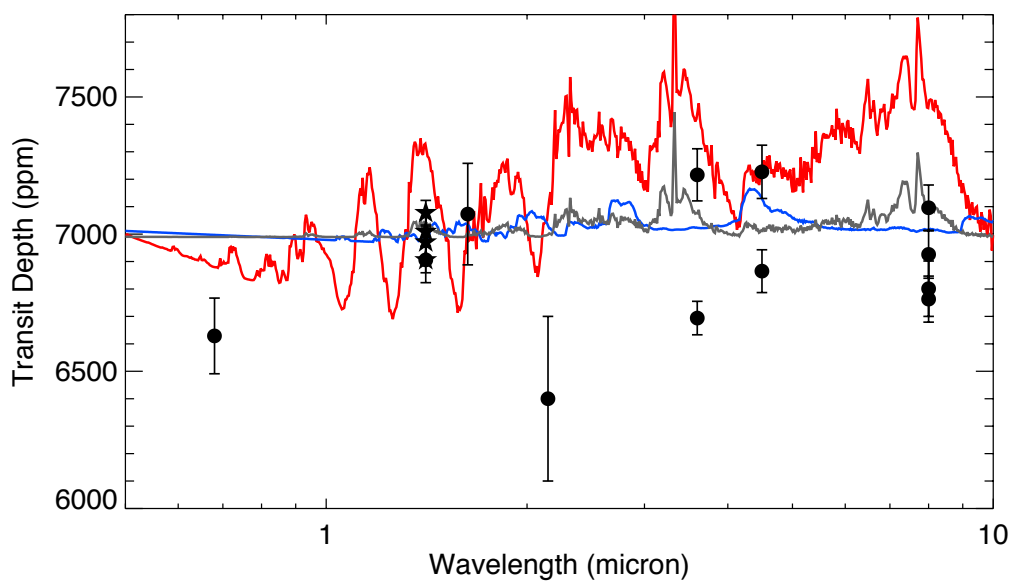

**Extended Data Figure 6. Comparison to published transit depths for GJ 436b.** Filled black circles show previously published transit depths[3,40,41,43,44]. The white-light transit depths from our WFC3 observations are overplotted as black stars. As we discuss in a previous study[3], the apparent variations in transit depth at different epochs could plausibly be explained by the occultation of active regions on the surface of the star. If correct, this would make broadband photometry collected at different epochs unreliable for the purpose of constraining the planet's transmission spectrum.

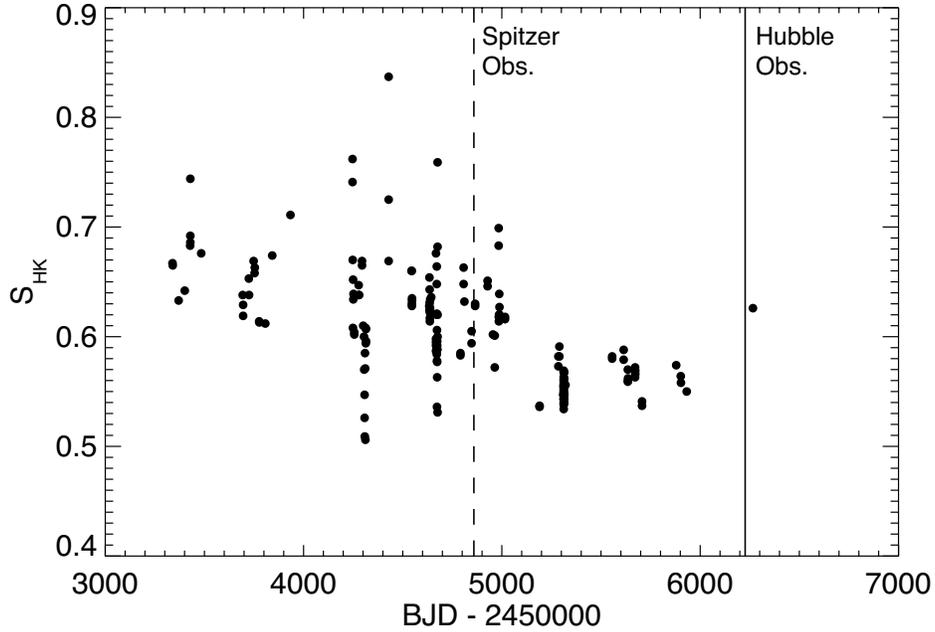

**Extended Data Figure 7. Stellar activity vs. time.** Filled black circles show the measured emission levels in the Ca II H & K line cores from Keck HIRES spectroscopy of GJ 436[3,45]; larger SHK values indicate increased stellar activity. Vertical lines mark the approximate dates of the six most recent Spitzer transit observations (dashed line), as well as the four HST transits presented in this paper (solid line).

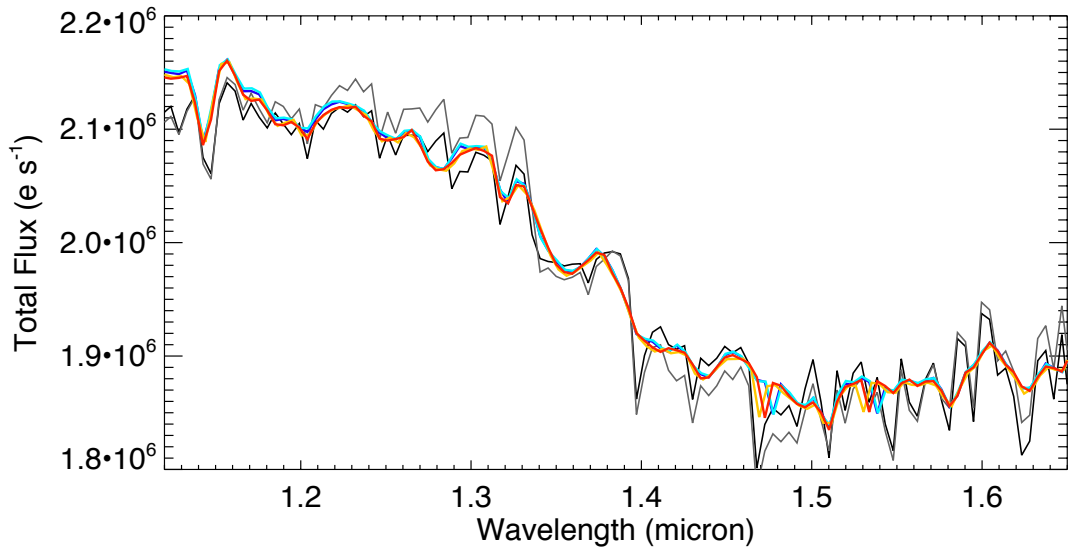

**Extended Data Figure 8. Averaged stellar spectrum vs. PHOENIX model atmospheres.** Spectra are averaged over each HST visit and then normalized using the sensitivity curve for that visit. These spectra are plotted as dark blue (Oct.), light blue (Nov.), yellow (Dec.), and red (Jan.) lines. For comparison we show two PHOENIX stellar atmosphere models with effective temperatures of 3500 K and log(g)=5.0 (black line) and 3350 K and log(g)=4.8 (gray line) binned to the same pixel resolution as our data. These data include an additional component of instrumental broadening that smooths out the sharp spectral features visible in the model spectra.

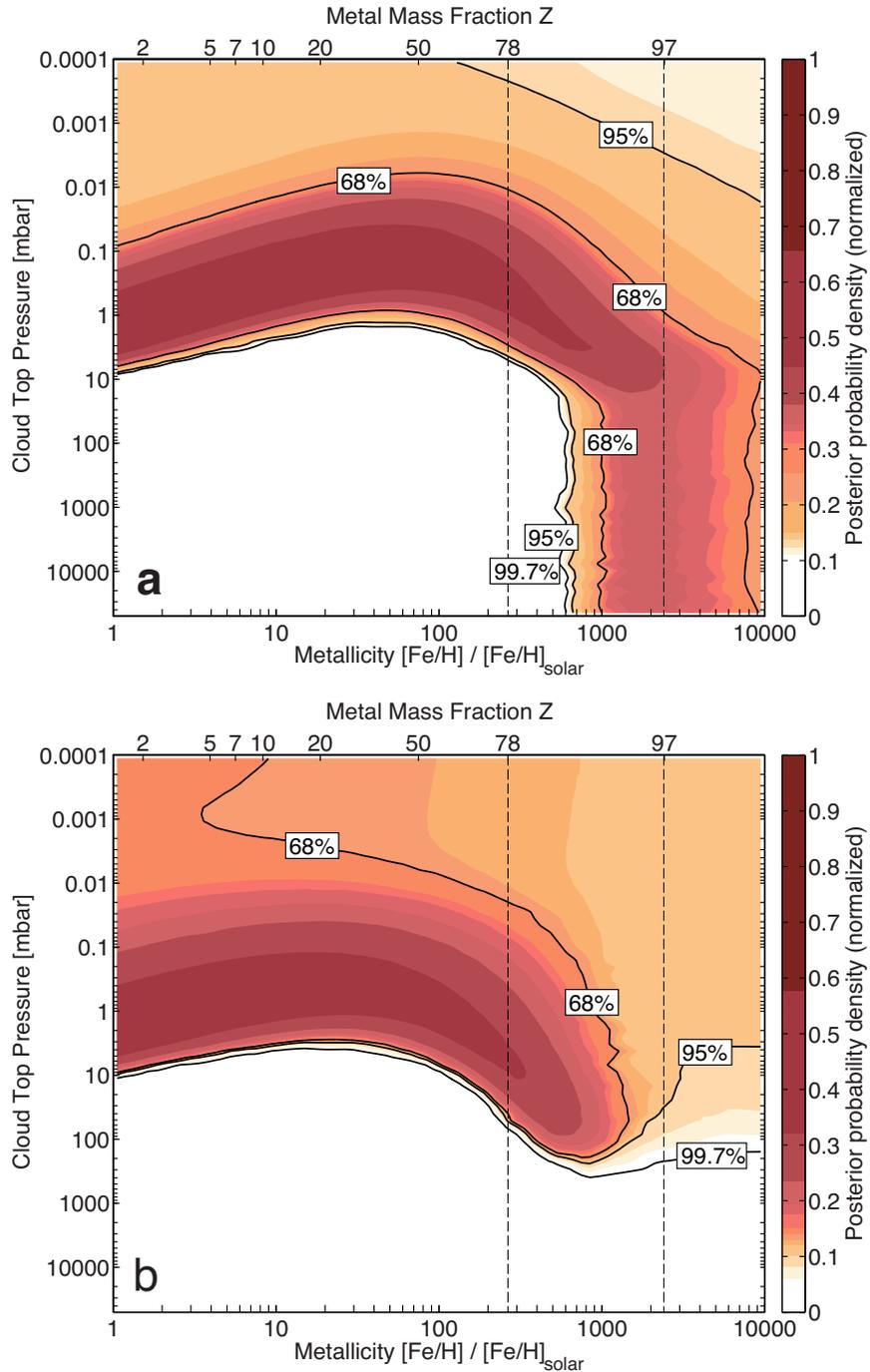

**Extended Data Figure 9. Joint constraints on cloud top pressure versus atmospheric metallicity** for an oxygen-rich (panel a; C/O = 0.3) and a carbon-rich (panel b; C/O = 1.0) atmosphere. The models shown in Fig. 3 assume a solar C/O ratio of 0.5. The colored shading indicates the normalized probability density as a function of cloud top pressure and metallicity derived from a variation of the Bayesian retrieval methods. We vary the amount of metals in the atmosphere (defined as elements heavier than H and He) linearly using the scaling factor shown on the lower x axis. The black contours show the 68%, 95%, and 99.7% Bayesian credible regions.

**Extended Data Table 1. Averaged Differential Transit Depths.**

| Wavelength | Depth (ppm) | Measurement Error (ppm) | Error from Limb Darkening (ppm) |
|---|---|---|---|
| 1.136 | 6966 | 60 | 7 |
| 1.155 | 6994 | 50 | 10 |
| 1.174 | 6924 | 40 | 12 |
| 1.193 | 6872 | 57 | 12 |
| 1.211 | 6968 | 39 | 17 |
| 1.230 | 7046 | 38 | 22 |
| 1.249 | 7036 | 39 | 20 |
| 1.268 | 6967 | 35 | 22 |
| 1.289 | 6989 | 35 | 24 |
| 1.306 | 7043 | 38 | 17 |
| 1.324 | 6989 | 38 | 20 |
| 1.343 | 7046 | 42 | 19 |
| 1.362 | 7057 | 37 | 25 |
| 1.381 | 7006 | 37 | 25 |
| 1.400 | 7036 | 50 | 27 |
| 1.419 | 7072 | 46 | 46 |
| 1.438 | 7030 | 42 | 46 |
| 1.456 | 7044 | 42 | 44 |
| 1.475 | 6948 | 39 | 44 |
| 1.494 | 7008 | 39 | 49 |
| 1.513 | 7057 | 40 | 55 |
| 1.532 | 7022 | 44 | 56 |
| 1.551 | 7018 | 40 | 46 |
| 1.570 | 7010 | 37 | 41 |
| 1.588 | 6959 | 40 | 41 |
| 1.607 | 6994 | 44 | 34 |
| 1.626 | 6984 | 44 | 30 |
| 1.645 | 6916 | 59 | 55 |